\documentclass[preprintnumbers,amsmath,prd,twocolumn,amssymb,floatfix,
superscriptaddress,nofootinbib]{revtex4}
\usepackage{graphicx}
\usepackage{epsfig}
\usepackage{bm}
\usepackage{amsfonts}
\def\be {\begin{equation}}
\def\ee {\end{equation}}
\def\ba {\begin{eqnarray}}
\def\ea {\end{eqnarray}}
\begin{document}

\title{Wormhole Thermodynamics at Apparent Horizons}

\author{\textbf{Mubasher Jamil}}
\email{mjamil@camp.nust.edu.pk} \affiliation{Center for Advanced
Mathematics and Physics, Campus of college of E\&ME, National University of
Sciences and
Technology,\\ Rawalpindi, 46000, Pakistan}

\author{\textbf{M. Akbar}}
\email{ak64bar@yahoo.com} \affiliation{Center for Advanced
Mathematics and Physics, Campus of college of E\&ME, National University of
Sciences and
Technology,\\ Rawalpindi, 46000, Pakistan}

\begin{abstract}\textbf{Abstract:} In this paper, we discuss the
thermodynamic properties of the evolving Lorentzian wormhole. For
the shape function $b(r) = r_{0}^2/r$, it is shown that the wormhole
spacetime admits two apparent horizons, the inner and the outer one.
The inner horizon expands while the outer contracts with the passage
of time. Corresponding to these horizons, we have three types of
wormholes, regular, extreme and the naked wormholes. Moreover, it is
shown that the Einstein field equations can be rewritten as a first
law of thermodynamics $dE=TdS+WdV$, at the apparent horizons of the
wormhole, where $E=\rho V$, $T = \kappa/2\pi$, $S=A/4G$,
$W=(\rho-P)/2$ and $V = \frac{4}{3}\pi \tilde{r}_{A+}^3$ are the
total matter energy, horizon temperature, wormhole entropy, work
density and the volume of the wormhole respectively.
\end{abstract}

\maketitle
\newpage
\section{Introduction}
Stephen Hawking showed \cite{hawking} that black holes emit thermal
radiation corresponding to a temperature proportional to surface
gravity and entropy proportional to the horizon area. The horizon
temperature and entropy obey a simple differential relationship
$-dE=TdS$, called the first law of black hole thermodynamics
\cite{bardeen}, where $E$ is the energy. Another significant
development was made by Jacobson \cite{jacobson} by deriving
Einstein field equations from the proportionality of entropy to the
horizon area together with the fundamental relation $\delta Q=TdS$,
where $\delta Q$ and $T$ are the energy flux and Unruh temperature
seen by an accelerated observer just inside the horizon.
\\
Padmanabhan \cite{Padmanabhan} made the major development by
launching a general formalism for the spherically symmetric black
hole spacetimes to understand the thermodynamics of horizons and
showed that the Einstein field equations evaluated at event horizon
can be expressed in the form of first law, $TdS = dE + PdV$, of
thermodynamics. Later on Padmanahban et al and others
\cite{paranjape,paranjape1} studied this approach for more
general spacetime geometries and in various gravity theories. \\
In the cosmological setup, Cai and his collaborators \cite{cai,cai1,
akbar1, cai2, akbar} made the major development by showing that the
Einstein field equations evaluated at the apparent horizon can also
be expressed as the first law $TdS = dE + WdV$ in various theories
of gravity. This connection between gravity and thermodynamics has
also been extended in the braneworld cosmology \cite{wang}. More
recently, using Clausius relation $\delta Q=TdS$, to the apparent
horizon of a FRW universe, Cai et al  are able to derive the
modified Friedman equation by employing quantum corrected
area-entropy formula \cite{rong}.  All these calculations indicate
that the thermal interpretation of gravity is to be generic, so we
have to investigate this relation for a more general spacetimes.
Hence, in this paper, we extend this approach for evolving
Lorentzian wormhole spacetime and showed by using the approach of
\cite{caiakbar} that the field equations of the wormhole geometry
can be expressed as a first law $TdS = dE + WdV$, at the apparent
horizon. We also study the dynamics of wormhole horizons. This paper
is organized as follows: In the second section, we discuss the
dynamics of wormhole horizons. In third section, we study the
thermal interpretation of the field equations at wormhole apparent
horizons. Finally, we present conclusion in the fourth section.
\section{Evolving Lorentzian wormhole}
A simple generalization of Morris-Thorne (MT) wormhole \cite{morris}
to the time dependent background is given by the evolving Lorentzian
wormhole \cite{cata} \be ds^2=-e^{2\Phi(t,r)}dt^2+a^2(t)\Big[
\frac{dr^2}{1-\frac{b(r)}{r}}+ r^2d\Omega_2^2 \Big]. \ee Here
$d\Omega_2^2\equiv d\theta^2+\sin^2\theta d\phi^2$ is the line
element of two dimensional unit sphere, $b(r)$ and $\Phi(t,r)$ are
the shape and potential functions respectively and $a(t)$ is the
scale factor of the universe. It is clear from the metric (1) that
 if both $b(r)\rightarrow0$, and $\Phi(t,r)\rightarrow0$, the above metric
reduces to the flat FRW metric. Furthermore, when $a(t)\rightarrow$
const and
 $\Phi(t,r)\rightarrow \Phi(r)$, it turns out the static MT wormhole
 \cite{morris}. If one takes $a(t)=e^{\chi t}$, the metric (1) represents
 an inflating Lorentzian wormhole \cite{roman}, where the arbitrary
 constant $\chi$ can be fixed by taking it a cosmological constant
 $\Lambda$.

In this paper, we shall use the ansatz $\Phi(t,r)=0$, and
$b(r)=r_0^2/r$, where $r_0$ is a finite radius of the wormhole's
throat. The former assumption is motivated from two facts; the
potential function must be finite quantity for all values of $r$,
for the static case. Moreover, it is also one of the solutions of
Einstein field equations for the wormhole spacetime
\cite{morris,peter}. This ansatz for the shape function clearly
obeys flare-out condition at the throat; $b(r_0)=r_0,$
$b^\prime(r_0)<1$ and $b(r)<r.$ It also fulfils another requirement
of asymptotic flatness. Thus, taking the above ansatz we have the
following evolving wormhole metric \be ds^2=-dt^2+a^2(t)\Big[
\frac{dr^2}{1-\frac{r_0^2}{r^2}}+r^2d\Omega_2^2 \Big]. \ee One can
rewrite the above metric in the spherical form \be
ds^2=h_{ab}dx^adx^b+ \tilde r^2d\Omega_2^2,\ \ \ a,b=0,1 \ee where
$\tilde{r}=a(t)r$ and $x^0=t$, $x^1=r$ and the two dimensional
metric \be
h_{ab}=\text{diag}\Big(-1,a(t)^2\Big(1-\frac{r_0^2}{r^2}\Big)^{-1}\Big)
.\ee Let us now consider the Einstein field equations \be
G_{\mu\nu}=8\pi GT_{\mu\nu},\ \ \ \mu,\nu=0,1,2,3 \ee where
$G_{\mu\nu}$ is the Einstein tensor and $T_{\mu\nu}$ is the
energy-momentum tensor of the matter fields. Due to the spherical
symmetry of the wormhole geometry (3), the stress energy tensor
$T_{\mu\nu}$ must be diagonal. The simplest one is that of a perfect
fluid described by a time dependent energy density $\rho(t)$ and
pressure $p(t)$ \be T^{\mu\nu}=(\rho+p)u^\mu u^\nu+pg^{\mu\nu}, \ee
where $u^\mu=(1,0,0,0)$ is the comoving four velocity of the fluid.
The energy conservation condition $T^{\mu\nu}_{;\nu}=0,$ admits
$\dot\rho+3H(\rho+p)=0$. Solving the Einstein field equation (5), in
the background of wormhole geometry (2), one can get the
Friedman-like field equations
\ba H^2-\frac{r_0^2}{3r^4a^2}&=&\frac{8\pi G}{3}\rho,\\
\dot H+\frac{r_0^2}{a^2r^4}&=&-4\pi G(\rho+p). \ea Here $H=\dot
a/a$, is the Hubble parameter and dot refers to the cosmic time
derivative. The apparent horizon of the wormhole geometry can be
evaluated by using relation $h^{ab}\partial_a\tilde r\partial _b
\tilde r=0$, which after simplification yields \be
H^2\tilde{r}_A^4-\tilde{r}_A^2+a^2r_0^2=0.\ee It can be seen from
Eq. (9) that when $r_0=0$, namely a flat FRW universe, the wormhole
apparent horizon $\tilde{r}_A$, has the same value as the Hubble
horizon, $\tilde r_A=1/H$. The Hubble parameter in terms of the
wormhole apparent radius is $H^2=1/\tilde{r}_A^2-a^2r_0^2/r^4$, and
its time derivative admits \be \dot
H=-\frac{\dot{\tilde{r}}_A}{H\tilde{r}_A^3}\Big(1-\frac{2a^2r_0^2}
{\tilde{r}_A^2}\Big)-\frac{a^2r_0^2}{\tilde{r}_A^4}. \ee The
apparent horizons of the wormhole metric (3) are the roots of the
Eq. (9) which admits \be
\tilde{r}_{A+}^2=\frac{1+\sqrt{1-4H^2a^2r_0^2}}{2H^2},\ \
\tilde{r}_{A-}^2=\frac{1-\sqrt{1-4H^2a^2r_0^2}}{2H^2}. \ee There are
three cases depending upon the roots. (a) Two distinct real roots
($1-4H^2a^2r_0^2>0$) refer as a usual wormhole geometry, (b) two
repeated real roots ($1-4H^2a^2r_0^2=0$) called as the `extreme
wormhole' geometry, (c) no real roots ($1-4H^2a^2r_0^2<0$) imply the
`naked wormhole'. If we assume that $0<r_0^2\ll1$, and neglecting
$O(r_0^4)$, it is possible to simplify the expressions for
$\tilde{r}_{A+}$ and $\tilde{r}_{A-}$, which give \be
\tilde{r}_{A+}^2=\frac{1}{H^2}-a^2r_0^2,\ \ \ \tilde{r}_{A-}^2=a^2
r_0^2. \ee It is evident from equation (12) that the outer apparent
horizon will contract while the inner horizon will expand with the
passage of time. These coincide at the extreme case
$\tilde{r}_{A+}=\tilde{r}_{A-}=\tilde{r}_{A}=1/\sqrt{2}H$. It is
also interesting to note that the wormhole horizons satisfy \be
\tilde{r}_{A+}^2+\tilde{r}_{A-}^2=\frac{1}{H^2},\ \
\tilde{r}_{A+}^2\tilde{r}_{A-}^2=\frac{a^2r_0^2}{H^2}. \ee In the
extreme case $\tilde{r}_{A-}=\tilde{r}_{A+}$, the quantity inside
the square root, $1-4H^2a^2r_0^2$, in (11) vanishes. In this case,
the wormhole parameters satisfy $ \dot{a}^2=\frac{1}{4r_0^2},$ which
upon integration gives $a(t)=\pm\frac{t}{2r_0},$ where the constant
of integration is assumed to be zero. It shows that the wormhole is
expanding uniformly if $a(t)>0$ and contracting if $a(t)<0$. Also
the naked wormhole is obtained if the discriminant
$1-4H^2a^2r_0^2<0$, which yields $ \dot a>\frac{1}{2r_0}.$
\section{Wormhole thermodynamics}
In this section, we discuss the thermodynamic properties of wormhole
at wormhole horizons. We assume that the entropy associated with the
outer horizon $\tilde{r}_{A+},$ of the wormhole is proportional to
the wormhole horizon area analogous to the black hole entropies. So
the entropy of the wormhole becomes \be
S=\frac{\pi\tilde{r}_{A+}^2}{G}=\frac{\pi}
{G}\Big(\frac{1+\sqrt{1-4H^2a^2r_0^2}}
{2H^2}\Big).
\ee The surface gravity is defined as \be \kappa
=\frac{1}{2\sqrt{-h}}\partial_a(\sqrt{-h}h^{ab}\partial_b\tilde r),
\ee where $h$ is the determinant of metric $h_{ab}$ (4). The direct
calculation of the surface gravity from Eq. (15) at the wormhole
horizon $\tilde{r}_{A+}$ yields \ba
\kappa&=&-\frac{\tilde{r}_{A+}}{2}\Big( \dot
H+2H^2-\frac{a^2r_0^2}{\tilde{r}_{A+}^4}  \Big),\\
&=&-\frac{1}{\tilde{r}_{A+}}\Big(1-\frac{\dot{\tilde{r}}_{A+}}
{2H\tilde{r}_{A+}}\Big)
\Big(1-\frac{2a^2r_0^2}{\tilde{r}_{A+}^2}\Big). \ea The factor
$-\frac{1}{\tilde{r}_{A+}}\Big(1-\frac{\dot{\tilde{r}}_{A+}}
{2H\tilde{r}_{A+}}\Big)$, in (17) is the general expression for the
surface gravity of FRW universe while the second factor
$\Big(1-\frac{2a^2r_0^2}{\tilde{r}_{A+}^2}\Big)$ has been appeared
due to wormhole geometry. This expression for the surface gravity
reduces to the expression for FRW universe when $r_0$ vanishes. The
horizon temperature $T=\kappa/2\pi$, of the wormhole is given by \ba
T &=&-\frac{1}{2\pi\tilde{r}_{A+}}\Big(1-\frac{\dot{\tilde{r}}_{A+}}
{2H\tilde{r}_{A+}}\Big)
\Big(1-\frac{2a^2r_0^2}{\tilde{r}_{A+}^2}\Big). \ea Now our purpose
is to rewrite the Friedman-like equation (8) for the wormhole as a
first law of thermodynamics. For this we follow the procedure
already developed in \cite{caiakbar} in which the field equations of
FRW universe have been expressed as a first law at the apparent
horizon. So first we evaluate the Friedman-like equation (8) at the
apparent horizon $\tilde{r}_{A+}$ which turns out \be
\Big(1-\frac{2a^2r_0^2}{\tilde{r}_{A+}^2}\Big)d\tilde{r}_{A+}=4\pi
GH\tilde{r}_{A+}^3(\rho+p)dt. \ee Now we multiply to the above
equation by a factor
$(1-\frac{\dot{\tilde{r}}_{A+}}{2H\tilde{r}_{A+}})$ and arranging
the terms, we get \begin{widetext} \be
-\frac{1}{2\pi\tilde{r}_{A+}}\Big(1-\frac{\dot{\tilde{r}}_{A+}}
{2H\tilde{r}_{A+}}\Big) \Big(1-\frac{2a^2r_0^2}
{\tilde{r}_{A+}^2}\Big)d\Big(\frac{\pi\tilde{r}_{A+}^2}{G}\Big)=-4\pi
H\tilde{r}_{A+}^3
\Big(1-\frac{\dot{\tilde{r}}_{A+}}{2H\tilde{r}_{A+}}\Big)(\rho+p)dt.\ee
\end{widetext} From equation (20), one can immediately
identify that the term on the left hand side is $TdS$, where
 $S=A/4G=4\pi \tilde{r}_{A+}^2/4G$, is the entropy of the
 wormhole. So the above equation
reduces to \be TdS=-4\pi H\tilde{r}_{A+}^3
\Big(1-\frac{\dot{\tilde{r}}_{A+}}{2H\tilde{r}_{A+}}\Big)(\rho+p)dt.\ee
Now we consider the total matter-energy $E=\rho V$, surrounded by
the apparent horizon $\tilde{r}_{A+}$ of the wormhole. Taking the
differential of $E$ and using the energy conservation relation, we
get \be dE=4\pi\tilde{r}_{A+}^2\rho
d\tilde{r}_{A+}-4\pi\tilde{r}_{A+}^3H(\rho+p)dt. \ee Using Eqs. (21)
and (22), we finally get \be dE=TdS+WdV,\ee where $W=(\rho-p)/2$ is
the work density. The expression (23) is the unified first law of
thermodynamics in the cosmological setup \cite{sean}. Summarizing,
by taking total matter energy density within the apparent horizon of
the wormhole, the Friedman-like equation can be expressed as a
thermodynamic identity. It is important to note that one can
associate the notions of temperature and entropy with the apparent
horizons of the wormhole analogous to the apparent horizon of FRW
universe. A trivial calculation shows that the similar thermal
interpretation of the field equations also hold at the inner horizon
$\tilde{r}_{A-}$ of the wormhole.

\section{Conclusion}
In this paper we have shown that the Friedman-like equation of the
wormhole geometry can be expressed as a form of first law,
$dE=TdS+WdV$, at the apparent horizon of the wormhole geometry. Here
$E=\rho V$ is the total energy of matter inside the horizons,
$W=(\rho-p)/2$ and $V$ are the work density and volume inside the
horizon respectively. In this approach, the entropy $S$ is assumed
to be quarter of the wormhole apparent horizon area. In the
cosmological setting, indeed, the Friedman equations of the FRW
universe can be expressed as a thermal identity at apparent horizon
\cite{cai,caiakbar}. Our results indicate that the dynamic apparent
horizon has a wider range of applications when one associates it
with the notions of temperature and entropy. It is also interesting
to investigate the properties of the wormhole geometry at extreme
case. In case $p=-\rho$, one gets the standard form of the first law
$dE=TdS-pdV$. In addition, we also studied the apparent horizons of
the evolving wormhole by assuming a particular shape function, $b(r)
= r_{0}^2/r$, of the wormhole. Under this assumption, it is found
that the evolving wormhole contains two apparent horizons. The outer
wormhole horizon will contract while the inner will expand. Both
horizon will coincide at the extreme case. One question may arise;
whether we can express the field equations of the wormhole geometry
in the extended theories of gravity. The work in this direction is
under investigation.

\end{document}